\documentclass[prl, twocolumn]{revtex4}

\usepackage{graphicx}
\usepackage{amsmath}
\usepackage{amssymb}

\begin{document}

\title{Testing Realism in Quantum Mechanics Through Charge Conservation}

\author{Victor Atanasov}
\address{Faculty of Physics, Sofia University, 5 blvd J. Bourchier, Sofia 1164, Bulgaria}
\email{vatanaso@phys.uni-sofia.bg}

\begin{abstract}
	
The universe is not locally realistic. Abandoning causality often appears more palatable than giving up on realism. This paper proposes two novel experimental setups to test realism’s failure using the conservation of electric charge. The first employs weak measurements of charge density in a double-slit interference setup. The second uses entangled charged particles in a Bell-type experiment, measuring electric field correlations to detect non-local charge distribution. Both leverage charge conservation to explore whether charge location remains indefinite until measured. These experiments offer a new perspective on quantum foundations, using charge as a probe to question whether the universe assigns definite properties only upon observation. A discussion on why charge is more appropriate for such experiments than mass is also present.
	
\end{abstract}

\maketitle

The experimental violation of Bell's inequalities\cite{bell} in the works of \cite{aspect, hensen, giustina} is the evidence that the world is not "locally realistic" \cite{englert}. Local realism assumes two key features: i.) physical properties of objects have definite values independent of measurement (realism); and ii.) no influence can travel faster than the speed of light (locality). If these assumptions hold true, then correlations between measurements of properties of two separated entangled systems should satisfy the Bell inequalities. Any experiment conducted so far finds these relations violated \cite{mermin}.

Locality is at the heart of causality. Therefore, a faster-than-light influence is constrained by relativity. The other view is that realism fails: particles don’t have definite properties until measured (a core feature of the Copenhagen interpretation) \cite{bohr}. {\it Does the universe "decide" outcomes only when observed? Does God play dice?}

It is astonishing that presently, the view towards the abandonment of causality seems more acceptable than the abandonment of realism. However, the lack of locality may not be enough \cite{weihs, zeilinger} and the failure of realism in the quantum world may turn out to be the cheaper solution. 

Here we suggest two independent methods to experimentally test the failure in realism.

The tests of realism we suggest circle around a relativistically conserved property giving objects their ability to interact, ergo exist - the electric charge. The conservation of electric charge is among the strictest conservation laws in Physics usually encoded in the continuity equation $\partial_t \rho + \partial_{\sigma} j^{\sigma}=0$, where $\rho$ is the charge density and $j^{\sigma}$ - the current density. 

\section{What we know so far}
The same equation describes the conservation of probability in quantum mechanics and therefore we can have a hint at the outcome of the proposed experiments. Note, $\rho$ would mean probability density $\propto |\psi|^2$ and $j^{\sigma}$ the probability current density. 

Consider the double-slit experiment. This experiment highlights that the the position of a particle, does not have definite values until measured. Imagine a source that emits particles, such as electrons or photons, directed toward a screen with two narrow slits. Beyond the screen is a detector that records where the particles land. Case 1.: no attempt is made to observe which slit the particles pass through, an interference pattern appears on the detector plane.  This suggests that each particle behaves as a wave, passing through both slits simultaneously and interfering with itself. Case 2.: if a device is added to detect which slit each particle goes through, the interference pattern disappears. Instead, the detector measures two distinct bands, as if the particles behave like classical objects with a definite path through one slit or the other. Meaning of Case 1.: since no measurement is made, the particle exists in a superposition, meaning it doesn’t have a definite position (i.e., it’s not fixed to one slit). The interference pattern emerges because the particle’s wave-like nature allows it to take all possible paths. Meaning of Case 2.: measuring which slit the particle passes through forces it to choose a definite path, collapsing its wave-like behavior into a particle-like state. This transition shows that the observable (position, in this case) only becomes definite upon measurement.

The double-slit experiment shows that physical observable, that is particle’s position, lack definiteness prior to measurement. It captures the essence of quantum mechanics, where the act of observation plays a critical role in determining the state of a system.

\bigskip

When a quantum charged particle, such as an electron, appears to be in two places at once—like in the double-slit experiment - it is described as being in a superposition of states. Its position is not definite but is instead represented by a wave function that assigns probabilities to finding the particle at different locations. A question then arises: {\it What happens to the electric charge in this situation?} The value of the charge does not change regardless of the particle’s quantum state. Whether the electron is in a superposition of positions or localized at a single point, its (total) electric charge remains definite and conserved: $-e$. 
However, the concept of being “in two places at once” affects how the charge is spatially distributed before a measurement is made.  

For a charged particle, the charge density $ \rho = -e \, |\psi|^2 $. If the particle is in a superposition, say between two locations A and B (e.g., through both slits in the double-slit experiment), the wave function might look like: $\psi = {(\psi_A + \psi_B)}/{\sqrt{2}}$, where $\psi_A$ and $\psi_B$ are wave functions localized at A and B, and the factor $1/\sqrt{2}$ ensures normalization. The charge density then becomes:

\[ \rho = -e \, |\psi|^2 = -e \, \frac{|\psi_A|^2 + |\psi_B|^2 + 2 \text{Re}(\psi_A^* \psi_B)}{2} \]

This shows that the charge density is smeared over the possible locations, with contributions at both A and B, and even interference terms depending on the overlap of the wave functions. However, the total charge remains conserved: $ \int \rho \, dV = -e $.

Before a measurement, the particle’s position and thus the location of its charge is indefinite. The charge is effectively smeared across the regions where the wave function has significant amplitude \cite{feynman}. When a measurement is made (e.g., detecting the electron at one of the slits or on a screen), the wave function collapses to a definite state. If the electron is found at location A, the charge density becomes concentrated at A, and the full charge $-e$ is localized there. The same applies if it’s found at B. Until that measurement, though, the charge’s location mirrors the uncertainty of the particle’s position.

This distribution can lead to observable effects, such as interference patterns in the electromagnetic field influenced by the charge, but the intrinsic charge itself is unaffected by the superposition. Finally, the charge of a quantum particle in a superposition of positions remains definite and conserved, but its spatial distribution is governed by the wave function until a measurement localizes it.

--------------------

Note, mass is not the same. When a quantum particle, such as an electron, exhibits behavior like being in two places at once, it raises an interesting question about its mass. When the electron’s position is indefinite due to superposition, does its mass get spread out or behave differently? Take the Schrödinger equation, which governs the evolution of the wave function for non-relativistic particles like electrons, mass appears as a constant parameter:

\[ i\hbar \frac{\partial \psi}{\partial t} = -\frac{\hbar^2}{2m} \nabla^2 \psi + V(\mathbf{r}) \psi \]

Here, $m$ is the particle’s mass, and it determines how quickly the wave function spreads over time. A larger mass results in slower spreading. In the superposition state of the double-slit experiment, the wave function is a combination of two parts - both components representing the particle passing through the A slit and the B slit use the "same mass". This indicates that mass is an intrinsic parameter of the electron itself, not something that changes or splits depending on the particle’s path. Mass is not treated like position or momentum, which are observables represented by operators with possible indefinite values in superposition. Mass, in this context, is a fixed parameter, not an observable subject to superposition itself.

In the experiment, the interference pattern depends on the electron’s wavelength, known as the de Broglie wavelength, given by: $\lambda = \frac{h}{\sqrt{2mK}}$, where $h$ is Planck’s constant and $p$ is momentum. For a given kinetic energy $K$, momentum is $p = \sqrt{2mK}$, therefore $m$ affects the wavelength and thus the spacing of the interference fringes. A particle with a different mass would produce a different pattern, but within the experiment, all electrons have the same mass, and it remains constant whether the wave function is in superposition or not.

\section{Set Up I: Interference}

\begin{figure}[ht]
	\begin{center}
		\includegraphics[scale=0.25]{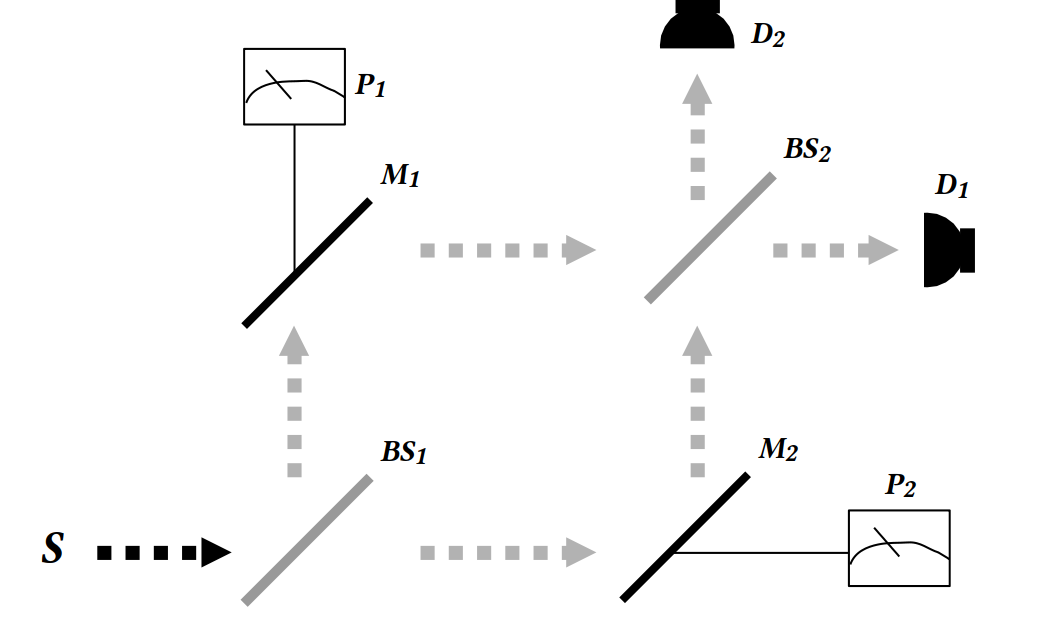}
		\caption{\label{fig1} An ion source (S) emits individual charged particles which go through a beam splitter (BS) 1. Now the quantum state of the charged particle is in a superposition between the two possible paths. Mirrors (M) guide the paths back to a beam splitter where they recombine to form an interference picture. Probes (P) perform weak measurement on the charge of the particle as it goes through the different paths. If probes measure a charge being present on the two possible paths while interference picture is maintained in the Detectors (D), the physical property ''charge'' exists independently from measurement, that is the case for realism.}
	\end{center}
\end{figure}

To design an experiment that tests realism in quantum mechanics using a particle’s charge in an interference setup, we need to address the core idea of realism: the notion that physical properties, such as the location of a particle’s charge, have definite values independent of measurement. Quantum mechanics challenges this by suggesting that properties like position and by extension - charge distribution, are undefined until measured, especially in superposition states. An interference experiment, such as the double-slit setup, is suited to testing this concept because it exhibits wave-like behavior (interference patterns) when the particle’s path is not measured, yet particle-like behavior (localized detection) when it is. 

The challenge is to incorporate the particle’s charge into this setup to probe whether its distribution has a definite location before measurement (realism) or exists in a superposition consistent with quantum mechanics.

The experimental design revolves around weak measurements in a double-slit setup \cite{aharonov}. Weak measurement is a technique that allows us to extract information about a quantum system without fully collapsing its wave function \cite{lundeen}. Unlike strong measurements, which force the system into an eigenstate of the measured observable, weak measurements provide subtle, averaged information about the charge density while preserving the interference pattern. This is ideal for testing whether the charge distribution matches the wave function’s probability density across multiple paths, rather than being localized to one path as realism would suggest. 

Let us use a source emitting charged particles, such as electrons (charge $q = -e $), one at a time. Each electron is prepared in a superposition state by passing through a double-slit apparatus. The electron’s wave function after the slits is: $\psi = {(\psi_A + \psi_B)}/{\sqrt{2}}$ representing a superposition of passing through slit A and slit B (or going along path A and B in a Mach-Zender type apparatus). The spatial probability density is $ |\psi|^2 $, and since the total charge $q$ is constant, the charge density is expected to be $q |\psi|^2 $.

Now, let us place sensitive charge detectors near each slit (along each path) to weakly measure the charge density (e.g. a small capacitor or electrometer). They interact weakly with the electron’s electric field if it passes through the respective slit (path). After passing the slits (paths), the electron proceeds to a detection screen, where its position is strongly measured, producing an interference pattern due to the superposition.

Suppose, the initial state is the above mentioned superposition. As the electron passes the slits (paths), its charge weakly interacts with the detectors. The interaction Hamiltonian is proportional to the charge density operator \( \hat{\rho}(\mathbf{r}) = q \delta(\mathbf{r} - \hat{\mathbf{r}}) \), where \( \hat{\mathbf{r}} \) is the position operator. The weak coupling shifts the pointer states of the detectors slightly, proportional to the local charge influence, with or without collapsing the electron’s wave function. We measure the electron’s final position on the screen (in detectors), selecting specific outcomes (e.g., a peak in the interference pattern).

The weak value of the charge density at a position \(\mathbf{r}\) (e.g., near slit A or slit B) is calculated as:
\[
\rho_w(\mathbf{r}) = \frac{\langle \phi | \hat{\rho}(\mathbf{r}) | \psi \rangle}{\langle \phi | \psi \rangle},
\]
where $ |\psi\rangle $ is the pre-selected state (superposition through both slits), and $ |\phi\rangle $ is the post-selected state (electron detected at a specific detector position). For a superposition state, $ \langle \hat{\rho}(\mathbf{r}) \rangle = q |\psi(\mathbf{r})|^2 $, but the weak value $\rho_w(\mathbf{r})$ can reveal the charge’s effective distribution conditioned on the final measurement.

If the electron is in a superposition, weak values of the charge density can be non-zero near both slits simultaneously, even when the electron is detected far from one slit (path). For example, \(\rho_w(\mathbf{r}_1)\) near slit A and \(\rho_w(\mathbf{r}_2)\) near slit B may both show fractional charge effects (e.g., values less than $q$ but greater than 0), summing to the total charge  $q$, reflecting the wave function’s distribution.

If realism holds, the charge is localized to one slit (either fully at slit A or slit B) before measurement. A weak measurement should then show the full charge $q$ at one slit and zero at the other, with no distribution across both.

By averaging the weak measurement outcomes over many trials, correlated with specific post-selected positions, we can map the effective charge distribution. Experimental evidence (e.g., from weak measurement studies in optics) suggests that weak values align with the quantum mechanical prediction, showing influence at both slits.

If the weak values indicate that the charge density is distributed across both slits (e.g., $\rho_w(\mathbf{r}_1) \approx q/2$, $\rho_w(\mathbf{r}_2) \approx q/2 $) in a manner consistent with $ q |\psi|^2 $, this suggests the charge did not have a definite location prior to the strong measurement at the screen (detector). This contradicts realism, supporting the quantum view that the charge distribution follows the wave function until measured.

An experiment using weak measurements of charge density in a double-slit interference setup effectively tests realism. By demonstrating that the charge appears distributed across both slits, consistent with the wave function’s probability density, before a strong measurement localizes it, the experiment supports quantum mechanics over realism. This leverages the electric charge’s conservation as a probe into the nature of quantum realism.

\section{Set Up II: Correlations}

In quantum mechanics, entanglement involves two (or more) particles whose states are correlated such that measuring one particle instantaneously determines the state of the other, regardless of distance. To test realism, we can use a Bell-type experiment, which compares the predictions of quantum mechanics against those of local hidden variable theories (a form of realism). The key is to leverage the charge of particles, ensuring the total charge is conserved while its spatial distribution remains probabilistic until measurement.

Since fundamental particles like electrons have a fixed charge (e.g., $-e$), we focus on the "position" of these charges rather than varying the charge itself. The spatial distribution of charge will follow the wave function’s probability density, and by entangling two particles, we can examine whether this distribution has definite values before measurement or exhibits quantum non-locality, challenging realism.

Now, consider two charged particles, each with charge $ q = -e $ (e.g., electrons), prepared in a spatially entangled state. The total charge is constant at $-2e$, but the positions of the charges are not fixed until measured. A suitable entangled state could be:
\[
 |\psi\rangle = \frac{1}{\sqrt{2}} \left( |\mathbf{a}_1, \mathbf{b}_2\rangle + |\mathbf{b}_1, \mathbf{a}_2\rangle \right) 
 \]

Here: $\mathbf{a}$ and $\mathbf{b}$ are two distinct spatial positions (e.g., $\mathbf{a} = (-d, 0, 0)$ and $\mathbf{b} = (d, 0, 0)$ along the x-axis, separated by distance $2d$). Subscripts 1 and 2 denote the two distinguishable particles (for simplicity, we assume distinguish-ability, though indistinguishable particles would require symmetrization). In this state particle 1 is found at $\mathbf{a}$, particle 2 is at $\mathbf{b}$, and vice versa. 

Before measurement, neither particle has a definite position; the charge distribution is a superposition governed by the wave function.
The charge density operator at a point \(\mathbf{r}\) is:
$
\hat{\rho}(\mathbf{r}) = q \delta(\mathbf{r} - \hat{\mathbf{r}}_1) + q \delta(\mathbf{r} - \hat{\mathbf{r}}_2), 
$
where $\hat{\mathbf{r}}_1$ and $\hat{\mathbf{r}}_2$ are the position operators of particles 1 and 2. The total charge, $\int \hat{\rho}(\mathbf{r}) d\mathbf{r} = 2q = -2e$ is conserved, but the spatial distribution depends on the entangled positions.

Directly measuring the charge density or particle positions collapses the wave function, fixing the charge distribution and making it difficult to test realism pre-measurement. {\it Instead, we measure the electric field produced by the charges}, which depends on their positions and thus reflects the probabilistic distribution. The electric field "operator" (since proportional to the position operator) at a point $\mathbf{r}$ is:
\[ 
\hat{\mathbf{E}}(\mathbf{r}) = \frac{q}{4\pi \epsilon_0} \left( \frac{\mathbf{r} - \hat{\mathbf{r}}_1}{|\mathbf{r} - \hat{\mathbf{r}}_1|^3} + \frac{\mathbf{r} - \hat{\mathbf{r}}_2}{|\mathbf{r} - \hat{\mathbf{r}}_2|^3} \right) 
\]

Since $\hat{\mathbf{r}}_1$ and $\hat{\mathbf{r}}_2$ are operators, $\hat{\mathbf{E}}(\mathbf{r})$ encodes the superposition of possible charge positions. It also is an observable since a linear combination of observables.

To adapt this to a Bell test, which requires "two-outcome" observables (e.g. spin projections; polarization), we define measurements based on the sign of the electric field components: 

i.) Alice’s measurements. 
At points $\mathbf{r}_{\alpha}$ and $\mathbf{r}_{\beta}$, measure the sign of a field component (e.g., $x$ or $y$-component).
$ A_1 = \text{sign}(E_x(\mathbf{r}_{\alpha})) $: +1 if positive, -1 if negative. $ A_2 = \text{sign}(E_y(\mathbf{r}_{\beta})) $.

ii.) Bob’s measurements. At points $\mathbf{r}_{\gamma}$ and $\mathbf{r}_{\delta}$, similarly: $ B_1 = \text{sign}(E_x(\mathbf{r}_{\gamma})) $ and $ B_2 = \text{sign}(E_y(\mathbf{r}_{\delta})) $.

The points $\mathbf{r}_{\alpha}, \mathbf{r}_{\beta}, \mathbf{r}_{\gamma}, \mathbf{r}_{\delta}$ are chosen strategically (e.g., near $\mathbf{a}$ or $\mathbf{b}$) to probe the field contributions from the entangled positions.

In a Clauser-Horne-Shimony-Holt (CHSH) Bell test, two parties (Alice and Bob) measure correlations between their observables. The CHSH parameter is \cite{chsh}:
$$ 
S = E(A_1 B_1) - E(A_1 B_2) + E(A_2 B_1) + E(A_2 B_2),
$$
where $ E(A_i B_j) = \langle A_i B_j \rangle $ is the expectation value of the product of outcomes (+1 or -1). Local realism predicts $ |S| \leq 2 $, while quantum mechanics can yield the Tsirelson's bound $ |S| \leq 2\sqrt{2} $.

The electric field at a point depends on whether particle 1 is at $\mathbf{a}$ or $\mathbf{b}$, and particle 2 at the complementary position. For example, if $\mathbf{r}_{\alpha}$ is near $\mathbf{a}$, $ E_x(\mathbf{r}_{\alpha}) $ is dominated by the particle at $\mathbf{a}$. In $ |\mathbf{a}_1, \mathbf{b}_2\rangle $, particle 1 contributes a field in one direction; in $ |\mathbf{b}_1, \mathbf{a}_2\rangle $, particle 2 contributes differently (if distinguishable).

By choosing measurement points and components (e.g., $\mathbf{r}_{\alpha}$ near $\mathbf{a}$, $\mathbf{r}_{\gamma}$ near $\mathbf{b}$), the field signs correlate due to entanglement. Exact values of $ E(A_i B_j) $ require computing $\langle \psi | \hat{E}_i \hat{E}_j | \psi \rangle $ and depend on the geometry, but entangled states often maximize correlations (e.g., $ S \approx 2\sqrt{2} $ with optimal settings, as in spin-based Bell tests).

Now, entangle two charged particles, e.g., a quantum dot system or ion trap, ensuring spatial separation (Alice and Bob at distant labs). Alice randomly selects $ A_1 $ or $ A_2 $, measures the field sign at $\mathbf{r}_{\alpha}$ or $\mathbf{r}_{\beta}$. Bob selects $ B_1 $ or $B_2 $, measures at $\mathbf{r}_{\gamma}$ or $\mathbf{r}_{\delta}$. The experiment goes as a repeat over many trials, recording outcomes and timestamps to ensure space-like separation and compute $ S $ from the correlation averages.

If experimentally $|S| > 2 $, the results violate Bell’s inequality, indicating that the charge distribution (via the electric field) lacks definite values before measurement, contradicting realism. Quantum mechanics predicts such violations for entangled states, while local realist theories cannot. If $|S| \leq 2 $, it supports realism, though loopholes (e.g., detection efficiency) must be addressed.

This experiment uses entangled charged particles and electric field measurements to test whether the spatial distribution of charge has definite values pre-measurement. By demonstrating non-local correlations that violate Bell’s inequality, it can provide evidence against realism, affirming that the charge distribution follows the probabilistic wave function until measured, consistent with quantum mechanics.

\begin{figure}[bt]
	\begin{center}
		\includegraphics[scale=0.24]{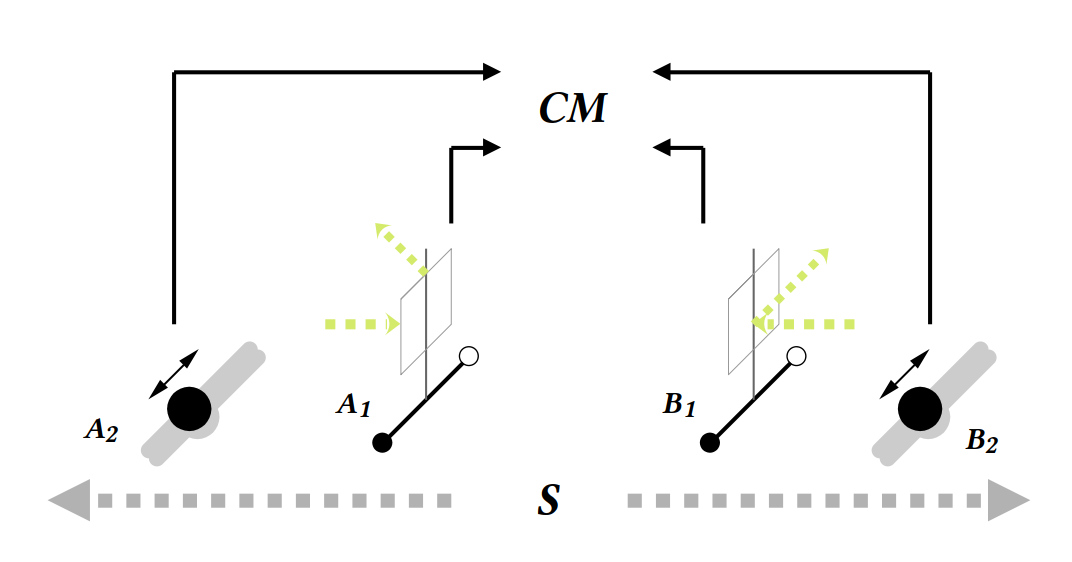}
		\caption{\label{fig2} A source (S) produces pairs of entangled charged particles traveling along distinct paths. Two sets of detectors at Alice's and Bob's sides: one (subindex 1) measures the presence (binary outcome $\pm 1$) of $x$-component to the electric field created by the particle (e.g. torsion balance type) and the second (subindex 2) - deals with the $y$-component (e.g. charge on a spring type - possibly MEMS). The signals from the detectors are fed into a coincidence monitor (CM). }
	\end{center}
\end{figure}

\bigskip

In conclusion, the presented experimental designs leverage the charge conservation and  entanglement to probe a foundational question in physics, namely the existence of properties/observables (realism) prior to measurement. The only necessity science imposes is to adjust our description of nature not to to our belief system but to the statistically significant results of an experiment. The present experimental bias is against realism in the quantum mechanical context but that requires further justification which is the scope of the proposed quite do-able tests. 

Useful discussions with the free version of Grok, an AI assistant by xAI, is acknowledged.

\end{document}